\documentclass{ws-ijmpe}
\usepackage[super,compress]{cite}

\begin{document}

\markboth{M. Thoennessen}{2015 Update of the Discoveries of Isotopes}

\catchline{}{}{}{}{}
%%%%%%%%%%%%%%%%%%%%%%%%%%%%%%%%%%%%%%%%%%%%%%%%%%%%%%%%%%%%%%%%%%%%

\title{2015 UPDATE OF THE DISCOVERIES OF NUCLIDES}

\author{\footnotesize M. THOENNESSEN}

\address{National Superconducting Cyclotron Laboratory and \\
Department of Physics \& Astronomy \\
Michigan State University\\
East Lansing, Michigan 48824, USA\\
thoennessen@nscl.msu.edu}

\maketitle

\begin{history}
\received{Day Month Year}
\revised{Day Month Year}
%\accepted{Day Month Year}
%\comby{(xxxxxxxxxx)}
\end{history}

\begin{abstract}
The 2015 update of the discovery of nuclide project is presented. Twenty new nuclides were observed for the first time in 2015. An overall review of all previous assignments was made in order to apply the discovery criteria consistently to all elements. In addition, a list of isotopes published so far only in conference proceedings or internal reports is included.
\end{abstract}


\ccode{PACS numbers: 21.10.-k, 29.87.+g}

%\tableofcontents

\section{Introduction}

This is the third update of the isotope discovery project which was originally published in a series of papers in Atomic Data and Nuclear Data Tables from 2009 through 2013 (see for example the first \cite{2009Gin01} and last \cite{2013Fry01} papers). Two summary papers were published in 2012 and 2013 in Nuclear Physics News \cite{2012Tho03} and Reports on Progress in Physics \cite{2013Tho02}, respectively, followed by annual updates in 2014 \cite{2014Tho01} and 2015 \cite{2015Tho01}. 

In preparation for the book ``The Discovery of Isotopes -- A complete Compilation'' to be published later this year \cite{2016Tho01} all discovery assignments were reevaluated to apply the criteria uniformly for all elements.

\section{New discoveries in 2015}
\label{New2015}

In 2015, the discoveries of twenty new nuclides were reported in refereed journals. Two proton-unbound resonances, one proton-rich nuclide, eleven neutron-rich nuclides, two $\alpha$-emitting uranium isotopes and four transuranium nuclides were discovered and are listed in Table \ref{2013Isotopes}.

\begin{table}[pt]
\tbl{New nuclides reported in 2015. The nuclides are listed with the first author, submission date, and reference of the publication, the laboratory where the experiment was performed, and the production method (PF = projectile fragmentation, FE = fusion evaporation, DI = deep inelastic reactions, SB = secondary beams). \label{2013Isotopes}}
{\begin{tabular}{@{}llrclc@{}} \toprule 
Nuclide(s) & First Author & Subm. Date & Ref. & Laboratory & Type \\ \colrule
$^{118}$Mo, $^{121}$Tc, $^{127}$Rh, $^{129}$Pd & G. Lorusso & 10/14/2014 & \refcite{2015Lor01} & RIKEN & PF \\
$^{132}$Ag, $^{134}$Cd, $^{136}$In, $^{137}$In & & & & & \\
$^{139}$Sn, $^{141}$Sb, $^{144}$Te  \vspace*{0.01cm}& & & & & \\
$^{216}$U & L. Ma & 4/9/2015 &  \refcite{2015Ma01} \vspace*{0.01cm} &  Lanzhou   & FE \\
$^{223}$Am, $^{229}$Am, $^{233}$Bk \vspace*{0.01cm} &  H. M. Devaraja  & 5/5/2015 &  \refcite{2015Dev01} &  GSI  & DI  \\
$^{59}$Ge \vspace*{0.01cm} & A. A. Ciemny & 6/16/2015 & \refcite{2015Cie01} & MSU & PF \\
$^{221}$U \vspace*{0.01cm} &   J. Khuyagbaatar & 7/14/2015 &  \refcite{2015Khu01}  &  GSI   & FE \\
$^{29}$Cl, $^{30}$Ar \vspace*{0.01cm} & I. Mukha & 8/6/2015 &  \refcite{2015Muk01}  &  GSI   & SB \\
$^{284}$Fl \vspace*{0.01cm} & V.K. Utyonkov & 8/31/2015 &  \refcite{2015Uty01}  &  Dubna   & FE \\
\botrule
\end{tabular}}
\end{table}

Lorusso {\it et al.} reported the discovery of eleven isotopes in the paper ``$\beta$-Decay Half-Lives of 110 Neutron-Rich Nuclei across the N= 82 Shell Gap: Implications for the Mechanism and Universality of the Astrophysical r Process''. 
A 345 MeV/nucleon $^{238}$U beam from the RIKEN cyclotron accelerator complex was incident on a beryllium target and ``After selection and identification, exotic nuclei were implanted at a rate of 50 ions/s in the stack of eight double-sided silicon strip detectors WAS3ABi, surrounded by the 84 high-purity germanium detectors of the EURICA array to detect $\gamma$ radiation from the
excited reaction products.''  \cite{2015Lor01} The table in the paper listing the measured half-lives included the newly identified isotopes $^{118}$Mo, $^{121}$Tc, $^{127}$Rh, $^{129}$Pd, $^{132}$Ag, $^{134}$Cd, $^{136}$In, $^{137}$In, $^{139}$Sn, $^{141}$Sb, and $^{144}$Te.

On April 9, 2015, Ma {\it et al.} submitted the paper describing the first observation of $^{216}$U entitled ``$\alpha$-decay properties of the new isotope $^{216}$U'' \cite{2015Ma01}. $^{216}$U was formed in the fusion evaporation reaction $^{180}$W($^{40}$Ar,4n)$^{216}$U with an 189.5 MeV $^{40}$Ar beam delivered from the Sector-Focusing Cyclotron of the Heavy Ion Research Facility in Lanzhou, China. The gas-filled recoil separator for Heavy Atoms and Nuclear Structure (SHANS) was used to separate evaporation residues which were implanted in a position-sensitive silicon strip detector (PSSD). $^{216}$U was identified by detecting correlated $\alpha$ particles in the PSSD or in a box of eight silicon detectors surrounding the PSSD in the backward direction. ``The granddaughter decay in two of the chains can be associated with $^{208}$Ra according to the literature values E$_\alpha$ = 7133(5) keV and T$_{1/2}$ = 1.3(2) s for the 0$^+$ ground state of $^{208}$Ra. Therefore, the $\alpha$ decay with E$_\alpha$ = 8384 $\pm$ 30 keV, T$_{1/2}$ = 4.72$^{+4.72}_{-1.57}$ ms can be recognized as belonging to $^{216}$U.''

Less than a month later, on May, 5 , 2015, Devajara {\it et al.} observed the decay of $^{216}$U in a deep inelastic multinucleon transfer reaction. They were apparently not aware of the results from Ma {\it et al.} which had not been published at the time of the submission. In addition to the observation of $^{216}$U, the paper entitled ``Observation of new neutron-deficient isotopes with Z $\le$ 92 in multinucleon transfer reactions'' \cite{2015Dev01} described the identification of the three new isotopes of  $^{223}$Am, $^{229}$Am, and $^{233}$Bk. The UNIversal Linear ACcelerator (UNILAC) at GSI was used to bombard layers of $^{248}$Cm oxide with a 270 MeV $^{48}$Ca beam. Target-like deep inelastic reaction products were separated and identified with the velocity filter SHIP by correlating implanted residues with subsequent $\alpha$-decays. ``[The] decay chain can be attributed to the new isotope $^{223}$Am. The decay sequence consists of an implanted recoil nucleus followed by a pileup $\alpha$ event with 17.4~MeV and three further $\alpha$ decays. The pileup of $\alpha_1$ and $\alpha_2$ is due to the expected short half-life of the daughter nucleus $^{219}$Np which is also a new isotope... We attributed [the] decay chain to the new isotope $^{233}$Bk. The observed half-lives of $^{233}$Bk and its daughter nucleus $^{229}$Am, which is also a new isotope, are well in agreement with WKB calculations using the measured $\alpha$ energies.'' The detection of $^{219}$Np mentioned in the quote cannot be credited with the discovery because neither its half-life nor its decay energy were measured.

Ciemny {\it et al.} discovered $^{59}$Ge at the National Superconducting Cyclotron Laboratory at Michigan State University in ``First observation of $^{59}$Ge" \cite{2015Cie01}. A 150 MeV/nucleon $^{78}$Kr beam bombarded a beryllium target and projectile fragments were separated and identified with the A900 separator. ``Four events can be identified as $^{59}$Ge on the basis of their position in the plot, also taking into account the `hole' corresponding to the unbound isotope $^{58}$Ga.''

Khuyagbaatar {\it et al.} described the discovery of $^{221}$U in the paper ``New Short-Lived Isotope $^{221}$U and the Mass Surface Near N = 126'' \cite{2015Khu01}. $^{50}$T beams between 231 and 255 MeV from the UNILAC at GSI bombarded $^{176}$YbF$_3$ targets at the gas-filled Trans Actinide Separator and Chemistry Apparatus (TASCA) at GSI to form $^{221}$U in the fusion evaporation reaction $^{176}$Y($^{50}$Ti,5n). ``The new isotope $^{221}$U was identified in $\alpha$-decay chains starting with E$_\alpha$ = 9.71(5)~MeV and T$_{1/2}$ = 0.66(14)~$\mu$s leading to known daughters.''

In the paper ``Observation and Spectroscopy of New Proton-Unbound Isotopes $^{30}$Ar and $^{29}$Cl: An Interplay of Prompt Two-Proton and Sequential Decay'' Mukha {\it et al.} identified $^{30}$Ar and $^{29}$Cl for the first time \cite{2015Muk01}. A secondary $^{31}$Ar was produced in projectile fragmentation from a 885 MeV/nucleon $^{36}$Ar beam from the SIS facility at GSI and separated with the FRS fragment separator. $^{30}$Ar was produced in one-neutron removal reactions on a secondary beryllium target located at the midplane of the FRS. In-flight decay products were measured with four large-area microstrip silicon detectors. ``Previously unknown isotopes $^{30}$Ar and $^{29}$Cl have been identified by measurement of the trajectories of their in-flight decay products $^{28}$S + p + p and $^{28}$S + p, respectively. The analysis of angular correlations of the fragments provided information on decay energies and the structure of the parent states. The ground states of $^{30}$Ar and $^{29}$Cl were found at 2.25$^{+0.15}_{-0.10}$ and 1.8$\pm$0.1 MeV above the two- and one-proton thresholds, respectively.''

The discovery of $^{284}$Fl was reported by Utyonkov {\it et al.} in ``Experiments on the synthesis of superheavy nuclei $^{284}$Fl and $^{285}$Fl in the $^{239,240}$Pu+$^{48}$Ca reactions'' \cite{2015Uty01}. It was produced in fusion evaporation reactions induced by $^{48}$Ca beams from the U400 cyclotron of the Flerov Laboratory of Nuclear Reactions, JINR in Dubna, Russia. Evaporation residues were implanted in an array of silicon detectors at the end of the Dubna Gas Filled Recoil Separator which also recorded subsequently emitted $\alpha$-particles and fission fragments. ``A new spontaneously fissioning (SF) isotope $^{284}$Fl was produced for the first time in the $^{240}$Pu + $^{48}$Ca (250 MeV) and $^{239}$Pu + $^{48}$Ca (245 MeV) reactions.''

\section{Changes of prior assignments}

The evaluation of the isotopes during the original discovery of the isotope project extended over a period of four years during which the criteria were adjusted and refined. Thus the criteria were not equally applied to all isotopes. In brief, a discovery of an isotope required for it to be (1) clearly identified, either through decay-curves and relationships to other known nuclides, particle or $\gamma$-ray spectra, or unique mass and element identification, and (2) published in a refereed journal. In order to avoid setting an arbitrary lifetime limit for the definition of the existence of a nuclide, particle-unbound nuclides with only short-lived resonance states were included. Isomers were not considered separate nuclides.

For the discovery of the first isotopes different criteria had to be applied because the concept of isotopes was only introduced by Soddy in 1913  \cite{1913Sod02}. Thus, in order to  be credited with the discovery of an isotope prior to 1913 at least one unique property had to be measured. After Rutherford established the radioactive decay law in 1900 \cite{1900Rut01} a fairly accurate half-life measurement was required to claim the discovery of a new isotope. The latter requirement changed the assignment for the discoveries of $^{227}$Ac and $^{219}$Rn. While Giesel observed their activities in 1902 \cite{1902Gie01} and 1903 \cite{1903Gie01}, respectively, he did not measure their half-lives. The half-life of ``a few seconds'' for $^{219}$Rn was reported by Debierne a month later \cite{1903Deb01} while the half-life of $^{227}$Ac was only determined in 1911 by M. Curie \cite{1911Cur01}.

Although J.J. Thomson was the first one to observe evidence for two substances with different atomic weight in a neon gas in 1913 \cite{1913Tho02,1913Tho01}, he did not accept the isotope concept that was developed by Soddy at the same time. As late as 1921 he argued that the heavier neon could be due to a hydride \cite{1921Tho03}. Thus the credit for the discovery of $^{20}$Ne and $^{22}$Ne was given to Aston, who stated at the 1913 meeting of the British Association for the Advancement of Science: ``...evidence has now been obtained that atmospheric neon is not homogeneous, but consists of a mixture of two elements of approximate atomic weights, 19.9 and 22.1 respectively... The two elements appear to be identical in all their properties except atomic weight.'' \cite{1913Ast01}

Originally the credit for the discovery of stable isotopes in mass spectrometers required the observation of two separate isotopes. However, Dempster demonstrated already in 1918 that his first mass spectrometer was capable of separating two neighboring isotopes \cite{1918Dem01}. Thus, his determinations of the masses of $^1$H (99.99\% abundance), $^{16}$O (99.76\%), $^{23}$Na (100\%), and $^{39}$K (93.26\%) can be considered the discovery of these isotopes.

The discovery assignments for stable isotopes of two additional elements were changed: $^{73}$Ge and $^{76}$Ge were changed from
Bainbridge \cite{1933Bai01} to Aston \cite{1931Ast05}, and $^{203}$Tl and $^{205}$Tl were changed from Aston \cite{1931Ast02} to Sch\"uler \cite{1931Sch01}.

The mass determination of fission fragments was initially very difficult and quite often ambiguous. The half-lives of many fission fragments were measured first without any firm mass assignment and credit for the discovery was given to the researchers who ultimately linked the activities to the correct mass. However, in the case of $^{92}$Sr G\"otte should be credited with the discovery although he did not determine its mass. G\"otte measured a 2.7~h strontium activity which he linked to an yttrium isotope with a half-life of 3.5~h \cite{1941Got01}. Unbeknownst to him Sagane {\it et al.} had assigned this activity to $^{92}$Y just one year earlier \cite{1940Sag01}.

Another difficulty was the assignment of the discovery of isotopes which were first reported in classified documents during the Second World War. Many fission fragments were identified within the Manhattan Project and the detailed results were only published in 1951 as part of the National Nuclear Energy Series \cite{1951Cor01}. However, a survey of the properties of the fission fragments had already been published in two simultaneous publications in the Journal of the American Chemical Society and Reviews of Modern Physics \cite{1946TPP01} quoting the still classified papers. Thus researchers at the time were aware of the results and credit for the discovery should be given to the initially classified work if it was included in the survey paper. Applying this criteria consistently for all fission fragments resulted in the reassignment of nine isotopes. It also should be mentioned that the first observation of fission fragments at Chalk River in Canada was published by Grummitt {\it et al.} about three months prior to the Manhattan Project survey, so they were credited with the discovery of all isotopes they correctly identified for the first time \cite{1946Gru01}.

The determination of the discovery of new fission fragments continued to be difficult after the War. In the early 1970s a technique was developed to measure decay properties of fragments from spontaneous fission sources  \cite{1970Wat01}. Again, exact mass identification was an issue and in many cases the observed $\gamma$-rays could not uniquely be assigned to a specific isotope. In order to be accepted as a discovery the mass had to be measured to within one mass unit and the $\gamma$-ray energies had to be confirmed later within the error bars and/or they had to be accepted in the level scheme in the Evaluated Nuclear Structure Data File ENSDF \cite{2008ENS01}. In order to be consistent, the assignment of four fission fragments ($^{108}$Tc \cite{1970Joh01}, $^{111}$Ru \cite{1970Joh01}, $^{120}$Cd \cite{1971Che01}, and $^{146}$La \cite{1974Aro01}) were changed.

A few other individual assignments were changed in order to adhere to various more detailed guidelines that had been adopted over the years.

For example, the assignment for $^{80}$As was changed from Ythier {\it et al.} \cite{1954Yth01} to Meads {\it et al.} \cite{1959Mea01} because the measured half-life of about 36~s was not within about a factor of two of the presently accepted value of 15.2$\pm$0.2~s \cite{2008ENS01}.

If the proton emitting daughter of a $\beta$-delayed proton emitter was not known at the time, the identification of the proton unbound states qualifies as the discovery of this isotope. Thus the discovery of $^{45}$V was reassigned to Jackson {\it et al.} who observed an excited state in the decay of the $\beta$-delayed proton emission of $^{45}$Cr \cite{1974Jac01}.

Although short-lived resonances of particle-unbound nuclides were included in the list of discovered isotopes, it is reasonable to set a lower limit on the lifetime. If the lifetime is shorter than about 10$^{-22}$ s which can be considered a characteristic nuclear timescale \cite{1960Gol01} it becomes questionable to classify the resonance as a nuclide. This lifetime corresponds to a width of about 6.6~MeV, so the observation of an 8~MeV wide resonance in $^5$H by Gornov {\it et al.} \cite{1987Gor01} should not qualify as the discovery of this isotope. Instead the credit was given to Korsheninnikov {\it et al.} who reported a resonance at 1.7$\pm$0.3~MeV with a width of 1.9$\pm$~MeV\cite{2001Kor01}.

In cases where the discovery of an isotope was reported essentially at the same time, the date of submission determined which research group received the credit for the discovery. However, if the first paper mentioned and acknowledged an unpublished report by the other group it is reasonable to give them the credit. Such a case was the discovery of $^{101}$Sn and $^{102}$Sn.  In the paper reporting the discovery of $^{100}$Sn submitted on April 27, 1993, Schneider {\it et al.} acknowledged the previous observation of $^{101}$Sn and $^{102}$Sn: ``Recently, an alternative production mechanism has been tested by using intermediate-energy projectile-fragmentation of 58 A$\cdot$MeV $^{112}$Sn at the LISE separator at GANIL. In this study, all the light $^{112}$Sn isotopes down to $^{101}$Sn could be identified'' \cite{1994Sch01}, referring to an article in the GANIL newsletter Nouvelles de GANIL by Lewitowicz {\it et al.} \cite{1993Lew01}. Lewitowicz {\it et al.} then submitted their results about six weeks later, on June 7, 1994, also reporting the observation of $^{100}$Sn \cite{1994Lew01}.

Another subjective criterion is the number of observed events required in projectile fragmentation reactions to qualify as the discovery of an isotope. There cannot be a common standard as it depends on the quality of the separation and selectivity of a given experimental setup. The confidence of the authors of these paper varied as the identification of only a few counts (sometimes a single count) were described as evidence, tentative, or the number of counts were listed in parentheses. In one instance, the selected criteria were not applied consistently for all  measured isotopes presented within one paper. In 1984, Hencheck {\it et al.} observed a few events for $^{77}$Y, $^{79}$Zr, $^{81}$Nb, $^{85}$Tc, $^{87}$Ru, $^{91}$Pd, and $^{93}$Ag, stating: ``Although these events all satisfy the stringent gating requirements mentioned above, we do not wish to state that such few events constitute proof that these nuclei were identified for the first time in this experiment as they could, at that level, be the result of `contamination' from neighboring peaks in the Z and Q spectra'' \cite{1994Hen01}. While in one of the first Atomic Data Nuclear Data Tables articles the observation of $^{93}$Ag was accepted \cite{2010Sch03}, later on the observation of the other isotopes was not accepted \cite{2012Nys01,2013Kat01}. Since Janas {\it et al.} demonstrated in 1999 that $^{81}$Nb and $^{85}$Tc were unbound \cite{1999Jan01} the observation of all isotopes mentioned by Hencheck {\it et al.} above should be questioned. Thus, the assignment of $^{93}$Ag was changed to the paper by Rykaczewski {\it et al.} who observed eight events of $^{93}$Ag in 1995 \cite{1995Ryk01}.

The assignments for the following five isotopes were changed for various different reasons: $^{189}$Au\cite{1955Smi01} (insufficient evidence for discovery), $^{189}$Hg\cite{1960Pof01} (original assignment too uncertain), $^{231}$Ra\cite{1985Hil01} (actually not measured in the previously assigned publication), $^{51}$Cl\cite{2009Tar01} (subsequently published arguments questioning the validity of the originally assigned publication), and $^{144}$Xe\cite{1989Bor02} (missed publication).

In a few additional cases a different publication from the same research group (most of the time with the same first author) was found to be more appropriate as the first paper reporting the discovery of new isotopes. One example is $^{271}$Ds, which had already been reassigned once before. Originally \cite{2013Tho01} credit was given to a 1998 review article by Hofmann \cite{1998Hof01} while last year \cite{2015Tho01} it was changed to an earlier paper by Armbruster, Hofmann, and Sobiczewski published in 1995 in Postepy Fizyki \cite{1995Arm01}. However, neither of the two previous assignments gave proper credit to all collaborators of the discovery experiment. The discovery of $^{271}$Ds was actually first reported in the November 1994 issue of the ``GSI - Nachrichten''\cite{1994Hof01}. It was then mentioned for the first time in the refereed literature in the discovery paper of element 111 which stated in the introduction: \cite{1995Hof02} ``In a recent experiment we produced the nucleus $^{269}$110 by the reaction $^{62}$Ni + $^{208}$Pb \cite{1995Hof01}. In a succeeding experiment we investigated the reaction $^{64}$Ni + $^{208}$Pb and observed the heavier isotope $^{271}$110 \cite{1994Hof01}.''  Because the author list of this paper is identical to the internal report it was credited with the discovery of $^{271}$Ds in order to acknowledge the contributions of all co-authors.

Finally, in some instances the laboratories where the experiments were performed had been listed incorrectly and have now been corrected.

\section{Status at the end of 2015}

\begin{figure}[pt] 
\centerline{\psfig{file=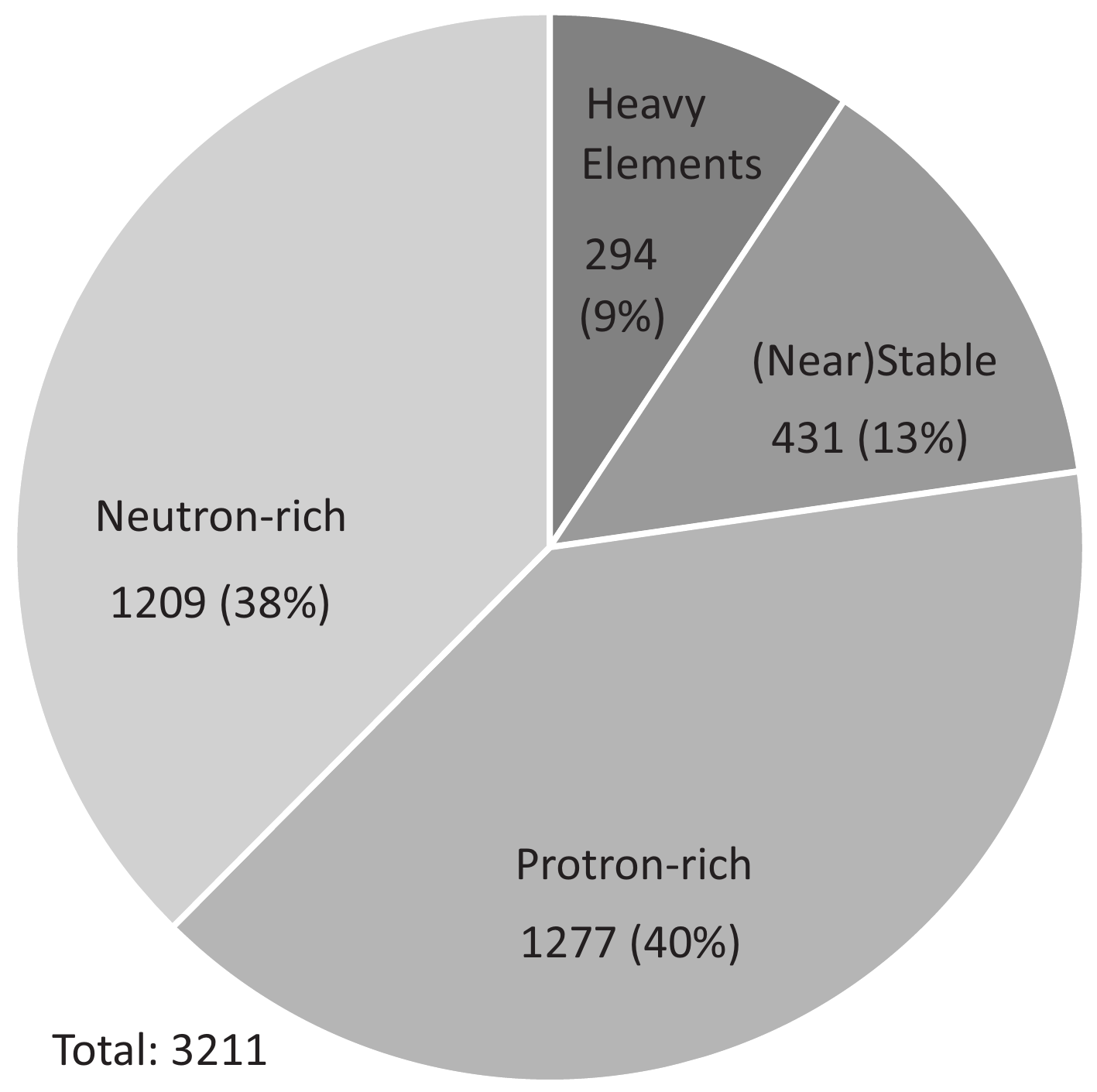,width=5.4cm}}
\caption{Total number of nuclides discovered for different areas of the chart of nuclides. \label{f:total} }
\vspace*{-0.7cm}
\end{figure}

The twenty new discoveries in 2015 increased the total number of observed isotopes to 3211. Figure \ref{f:total} shows the relative fractions of the broad areas of the nuclear chart, (near)stable, proton-rich, neutron-rich, and the region of the heavy elements. Although the rate of discoveries for neutron-rich isotopes continues to be significantly larger than for proton-rich isotopes, overall the latter still corresponds to the largest fraction. The ten-year average rate for neutron-rich isotopes reached an all time high of 23 isotopes per year, while the rate for proton-rich isotopes dropped below 3 isotopes per year corresponding to the lowest value since 1937.

The average discovery rate of heavy (transuranium) elements was about four per year. This rate has been fairly constant over the years fluctuating between two and eight per year since the 1950s.

It can be argued that the last (near)stable isotope discovered was $^{222}$Fr which is located between $^{221}$Fr and $^{223}$Fr of the neptunium and actinium radioactive decay chains, respectively. $^{222}$Fr was first observed in 1975 by Westgaard {\it et al.} at CERN \cite{1975Wes01}.

\begin{table}[b] 
\tbl{Top ten countries where the most nuclides were discovered. The total number of nuclides are listed together with first and most recent year of a discovery. \label{countries}}
{\begin{tabular}{@{}rlrcc@{}} \toprule
Rank & Country & Number & First year & Recent year \\ \colrule
1	&	 USA 		&	1327	&	1907	&	2015	\\
2	&	 Germany 	&	558	&	1898	&	2015	\\
3	&	 UK 		&	299	&	1900	&	1994	\\
4	&	 Russia 	&	247	&	1957	&	2015	\\
5	&	 France 	&	217	&	1896	&	2005	\\
6	&	 Japan 	&	139	&	1938	&	2015	\\
7	&	 Switzerland 	&	125	&	1934	&	2009	\\
8	&	 Canada 	&	66	&	1900	&	1998	\\
9	&	 Sweden 	&	59	&	1945	&	1993	\\
10	&	 Finland 	&	39	&	1961	&	2014	\\
\botrule
\end{tabular}}

\vspace*{0.8cm}

\tbl{Top ten laboratories where the most nuclides were discovered. The total number of nuclides are listed together with the country of the laboratory and the first and most recent year of a discovery. \label{labs}}
{\begin{tabular}{@{}rllrcc@{}} \toprule
Rank & Laboratory & Country & Number & First year & Recent year \\ \colrule
1	&	 Berkeley 	& USA		&	638	&	1928	&	2010	\\
2	&	 GSI	 	& Germany	&	438	&	1977	&	2015	\\
3	&	 Dubna 	& Russia	&	221	&	1957	&	2015	\\
4	&	 Cambridge 	& UK		&	218	&	1913	&	1940	\\
5	&	 CERN 	& Switzerland&	115	&	1965	&	2009	\\
6	&	 Argonne 	& USA		&	111	&	1947	&	2012	\\
7	&	 RIKEN 	& Japan	&	97	&	1938	&	2015	\\
8	&	 GANIL 	& France	&	86	&	1985	&	2005	\\
9	&	 Oak Ridge 	& USA		&	79	&	1946	&	2006	\\
10	&	 Michigan State 	& USA &	73	&	1967	&	2015	\\
\botrule
\end{tabular}}
\end{table}

The overall review of all discoveries resulted in a few changes of the various statistics previously presented. The 3211 isotopes were reported by 900 different first authors in 1531 papers and a total of 3598 different coauthors. Tables \ref{countries} and \ref{labs} list the top ten countries and laboratories, respectively, where isotopes were discovered. Further statistics can be found on the discovery project website \cite{2011Tho03}.

There were no major changes relative to last year's update \cite{2015Tho01}. The eleven new isotopes measured by Lorusso {\it et al.} at RIKEN  \cite{2015Lor01} gave Japan sole possession of sixth place in the country list, moving Switzerland to seventh place. The reassignment of the 1959 discovery of $^{142}$Pm and $^{142}$Sm by Gratot {\it et al.} \cite{1959Gra01} to Saclay moved Orsay out of the top-ten laboratory list and gave Michigan State University sole position of tenth place.

\section{Discoveries not yet published in refereed journals}

Table \ref{reports} lists all isotopes that have as of yet only been reported in conference proceedings or internal reports. None of the isotopes listed in last year's update \cite{2015Tho01} have been published in refereed publications this year. A few additional isotopes had not been included in the corresponding table last year. The two proton-rich $\alpha$-emitters $^{178}$Pb and $^{215}$U had been reported at PROCON 2003 \cite{2003Bat01}, and in the 2013 RIKEN Accelerator Report \cite{2014Wak02}, respectively. Also, $^{143}$Er was populated by the proton emitter $^{144}$Tm \cite{2005Grz01,2005Ryk01,2005Bin01} and $^{230}$At was identified as the parent of $^{230}$Pu decaying by electron capture in the decay chain of $^{234}$Bk \cite{2003Mor02,2003Mor01,2010Kaj01}. $^{230}$At and $^{232}$Rn had been listed in the 2013 update \cite{2014Tho01} but were inadvertently omitted last year. 

A large number of nuclides that had been discovered at RIBF in RIKEN, Japan, and presented at various conferences in 2013 and 2014 were published in conference proceedings in 2015. In total there are now over 70 isotopes produced by projectile fragmentation and projectile fission at RIBF that should be published in refereed journals in the near future. With the exception of $^{215}$U and maybe $^{230}$Am and $^{234}$Bk it is unlikely that any of the other observations listed in the table will appear in the refereed literature because they have been reported more than five years ago.

\begin{table}[t]
\tbl{Nuclides only reported in proceedings or internal reports until the end of 2015. The nuclide, first author, reference and year of proceeding or report are listed. \label{reports}}
{\begin{tabular}{@{}llrr@{}} \toprule
\parbox[t]{6.8cm}{\raggedright Nuclide(s) } & \parbox[t]{2.3cm}{\raggedright First Author} & Ref. & Year \\ \colrule

$^{20}$B		&	 F. M. Marqu\'es 	&	\refcite{2015Mar01}	&	2015	 \\ 
$^{21}$C		&	 S. Leblond 	&	\refcite{2015Leb01}	&	2015	 \\ 
$^{86}$Zn, $^{88}$Ga, $^{89}$Ga, $^{91}$Ge, $^{93}$As, $^{94}$As, $^{96}$Se, $^{97}$Se	&	 Y. Shimizu 	&	\refcite{2015Shi01}	&	2015	 \\ 
$^{99}$Br, $^{100}$Br & & & \\
$^{81}$Mo, $^{82}$Mo, $^{85}$Ru, $^{86}$Ru	&	 H. Suzuki	&	\refcite{2013Suz01}	&	2013	 \\ 
$^{90}$Pd, $^{92}$Ag, $^{94}$Cd	&	 I. Celikovic 	&	\refcite{2013Cel01},\refcite{2014Cel01}	&	2013/14	 \\ 
$^{96}$In, $^{98}$Sn, $^{104}$Te	&	  I. Celikovic 	&	\refcite{2013Cel01}	&	2013	 \\ 
$^{122}$Tc, $^{125}$Ru, $^{130}$Pd, $^{131}$Pd, $^{140}$Sn, $^{142}$Sb, $^{146}$I	&	 Y. Shimizu 	&	\refcite{2014Shi02}	&	2014	 \\ 
$^{153}$Ba, $^{154}$La, $^{155}$La, $^{156}$Ce, $^{157}$Ce, $^{156}$Pr$^a$, $^{157}$Pr, 	&	 D. Kameda 	&	\refcite{2013Kam01}	&	2013	 \\
$^{158}$Pr, $^{159}$Pr, $^{160}$Pr, $^{162}$Nd, $^{164}$Pm, $^{166}$Sm & & & \\
$^{155}$Ba, $^{157}$La, $^{159}$Ce, $^{161}$Pr, $^{164}$Nd, $^{166}$Pm, $^{168}$Sm,  	&	 N. Fukuda 	&	\refcite{2015Fuk01}	&	2015	 \\
$^{170}$Eu, $^{172}$Gd, $^{173}$Gd, $^{175}$Tb, $^{177}$Dy, $^{178}$Ho, $^{179}$Ho,& & & \\
 $^{180}$Er, $^{181}$Er, $^{182}$Tm, $^{183}$Tm & & & \\
	$^{156}$La, $^{158}$Ce, $^{163}$Nd, $^{165}$Pm, $^{167}$Sm, $^{169}$Eu, $^{171}$Gd,	&	 D. Kameda 	&	\refcite{2014Kam02}	&	2014	 \\
$^{173}$Tb, $^{174}$Tb, $^{175}$Dy, $^{176}$Dy, $^{177}$Ho, $^{179}$Er & & & \\
$^{126}$Nd, $^{136}$Gd, $^{138}$Tb, $^{143}$Ho$^b$, $^{150}$Yb, $^{153}$Hf	&	 G. A. Souliotis 	&	\refcite{2000Sou01}	&	2000	 \\
	$^{143}$Er, $^{144}$Tm	&	 R. Grzywacz 	&	\refcite{2005Grz01}	&	2005	 \\
	& K. Rykaczewski & \refcite{2005Ryk01} & 2005 \\
	& C. R. Bingham & \refcite{2005Bin01} & 2005 \\
$^{178}$Pb	&	 J. C. Batchelder 	&	\refcite{2003Bat01}	&	2003	 \\
$^{230}$At, $^{232}$Rn	&	 J. Benlliure 	&	\refcite{2010Ben02}	&	2010	 \\
$^{215}$U	&	 Y. Wakabayashi 	&	\refcite{2014Wak02},\refcite{2015Wak01}	&	2014/15	 \\
$^{230}$Am, $^{234}$Bk &	K. Morita & \refcite{2003Mor02} & 2003 \\
	&  K. Morimoto 	&	\refcite{2003Mor01}	&	2003	 \\
	& D. Kaji & \refcite{2010Kaj01}&	2010	 \\
$^{234}$Cm$^c$, $^{235}$Cm	&	 J. Khuyagbaatar 	&	\refcite{2007Khu01}	&	2007	 \\
$^{252}$Bk, $^{253}$Bk	&	 S. A. Kreek 	&	\refcite{1992Kre01}	&	1992	 \\
$^{262}$No 	&	 R. W. Lougheed 	&	\refcite{1988Lou01},\refcite{1989Lou01}	&	1988/89	 \\
	&	 E. K. Hulet 	&	\refcite{1989Hul01}	&	1989	 \\
$^{261}$Lr, $^{262}$Lr	&	 R. W. Lougheed 	&	\refcite{1987Lou01}	&	1987	 \\
	&	 E. K. Hulet 	&	\refcite{1989Hul01}	&	1989	 \\
	&	 R. A. Henderson 	&	\refcite{1991Hen01}	&	1991	 \\
$^{255}$Db	&	 G. N. Flerov	&	\refcite{1976Fle01}	&	1976	 \\
\botrule
\vspace*{-0.2cm} & & & \\
$^a$ also published in ref. \refcite{1996Cza01} and \refcite{1997Ber02} \\
$^b$ also published in ref. \refcite{2003Sew02} \\
$^c$ also published in ref. \refcite{2010Kaj01} and \refcite{2002Cag01}
\end{tabular}}
\end{table}

\section{Outlook for 2016}
Thirteen of the twenty isotopes discovered in 2015 were mentioned in last year's outlook. $^{216}$U  and $^{284}$Fl had been presented by Wakabayashi \cite{2014Wak01} and Rykaczewski \cite{2014Ryk01}, respectively, at the second Conference on Advances in Radioactive Isotope Science (ARIS). The eleven isotopes reported by Lorusso {\it et al.} \cite{2015Lor01} were included in the twenty-five nuclides presented by Shimizu \cite{2014Shi01} at the 4$^{th}$ Joint Meeting of the APS Division of Nuclear Physics and the Physical Society of Japan. Both conferences were held in 2014. It should be mentioned that $^{216}$U was first published by Ma {\it et al.}  \cite{2015Ma01} and shortly after by Devajara {\it et al.}  \cite{2015Dev01} (see Section \ref{New2015}). The results by Wakabayashi reported at ARIS have not yet been published in a refereed journal.

The number of isotopes predicted to be discovered last year (30) was somewhat higher than the actual number of discoveries (20). However, with the exception of $^{280}$Ds and seven  ($^{116}$Nb, $^{138}$In, $^{143}$Sb, $^{145}$Te, $^{147}$I, $^{149}$Xe, and $^{154}$Ba) of the isotopes presented by Shimizu at the APS/JPS meeting, all  other isotopes listed in last year's outlook were reported in conference proceedings or internal reports. 

In the next couple of years the majority of the over seventy neutron-rich isotopes measured at RIKEN and listed in Table \ref{reports} should be published in refereed journals. This should move RIKEN to fifth place in the list of top laboratories overtaking Argonne and CERN.

\section{Summary}

During the last year the assignments for the more than 3000 isotope discoveries were reviewed in preparation for the book `` The Discovery of Isotopes '' which will be published later this year \cite{2016Tho01}. A common set of criteria was applied to all isotopes which resulted in about ninety changes to the first author, reference or laboratory of the discoveries. Although no major reassignments are anticipated in the future, continued input, feedback, and comments from researchers are encouraged.

\section*{Acknowledgements}

I would like to thank Ute Thoennessen for carefully proofreading the manuscript. Support of the National Science Foundation under grant No. PHY11-02511 is gratefully acknowledged.

%\vfill
%\newpage

\bibliographystyle{ws-ijmpe}
%\bibliography{test2}
\bibliography{../springer-book-recover/tex-file/isotope-references-etal}

\end{document}